\documentclass[prd,preprint,superscriptaddress,showpacs,byrevtex]{revtex4}
\usepackage{epsfig}

\newcommand{\be}{\begin{equation}}
\newcommand{\ee}{\end{equation}}
\newcommand{\bea}{\begin{eqnarray}}
\newcommand{\eea}{\end{eqnarray}}
\def\simgt{\rlap{\lower 3.5 pt \hbox{$\mathchar \sim$}} \raise 1pt
Ê \hbox {$>$}}

\begin{document}

\title{ String Theory at LHC Using Higgs Production and Decay From String Balls}

\author{Gouranga C. Nayak} \email{nayak@physics.arizona.edu}
\affiliation{ Department of Physics, University of Arizona, Tucson, AZ 85721 USA
}


\begin{abstract}

We study Higgs production and decay from TeV scale string balls in $pp$ collisions at $\sqrt{s}$ = 14 TeV at LHC.
We present the results of total cross section of diphotons, invariant mass distribution of the diphotons and $p_T$
distribution of the diphotons and $ZZ$ pairs from Higgs from string balls at LHC.
We find that the invariant mass distribution of diphotons from Higgs from string balls
is not very sensitive to the increase in diphoton invariant mass.
We find that for string mass scale $M_s$=2.5 TeV, which is the lower limit of the
string mass scale set by the recent CMS collaboration at LHC, the
$\frac{d\sigma}{dp_T}$ of high $p_T$ ($\ge$ 450 GeV) diphotons and $ZZ$ pairs produced from Higgs from string balls
is larger than that from standard model Higgs. Hence in the
absence of black hole production at LHC an enhancement of high $p_T$ diphotons and $ZZ$
pairs at LHC can be useful signatures for string theory at LHC. Since the matrix element for Higgs production in parton-parton
collisions via string regge excitations is not available we compute $\frac{d\sigma}{dp_T}$ of photon production from string regge excitations
and make a comparison with that from string balls at LHC. We find that for string mass scale
$M_s$ = 2.5 TeV the $\frac{d\sigma_{photon}}{dp_T}$ from string
regge excitations is larger than that from string balls and black holes at LHC.

\end{abstract}
\maketitle

\newpage

One of the primary goal of the LHC experiment at CERN is to search for Higgs boson which is the last
missing particle of the standard model of particle physics. Higgs is responsible for generating particle masses.
It is now generally accepted that the scale of quantum gravity {\it could be} as low as one TeV
\cite{folks,gram}. In the
presence of extra dimensions, the string mass scale $M_s$ and the
Planck mass $M_P$ could be around $\sim$ TeV and we may expect to discover exotic physics at LHC such as
mini-black holes and TeV scale string balls. Pragmatically we can only search for Higgs, extra dimensions,
black holes and string balls at LHC in whatever mass range and energy range currently available at experiments.
While detection of Higgs makes the standard model complete, the production of mini-black holes will establish the existence of
extra dimensions and could possibly provide an experimental test of the quantum theory of gravity.
Similarly, detection of string  balls will provide experimental test of string theory for the first time. Hence if
Higgs, black holes and string balls are discovered at LHC then we will be propelled into the twenty first
century, as our understanding of quantum gravity and perhaps string theory is revolutionized.
Once black holes and string balls are produced at LHC they will
quickly evaporate to standard model particles including Higgs. Hence Higgs boson
production from black holes and string balls at LHC can play an important role in
distinguishing between Higgs from standard model electroweak physics and from quantum gravity
and/or from string theory. For example, if we do not find Higgs from standard model processes at LHC
but find from the decay of black holes and/or string balls then Higgs discovery at LHC may be related
to gravity and/or string theory rather than to the standard model of electroweak theory.
In this paper we make a detail analysis of the Higgs production
from black holes and string balls at LHC and make a comparison with the Higgs production
from standard model processes.

Recently, string theory has given convincing microscopic calculation for black hole evaporation
\cite{susskind,allstring}. String theory predicts that a black hole has formed at several times
the Planck scale and any thing smaller will dissolve into some thing
known as string ball \cite{dimo}. A string ball is a highly
excited long string which emits massless (and massive) particles at Hagedorn
temperature with thermal spectrum \cite{amati,canada}. For general relativistic
description of the back hole to be valid, the black hole mass $M_{BH}$ has to be
larger than the Planck mass $M_P$. In string theory the string ball mass $M_{SB}$
is larger than the string mass scale $M_s$. Typically
\bea
 M_s < M_P < \frac{M_s}{g_s^2},~~~~~~~~~~M_s << M_{SB} << \frac{M_s}{g_s^2},~~~~~~~~~\frac{M_s}{g_s^2} << M_{BH}
\label{scale}
\eea
where $g_s$ is the string coupling which can be less than 1 for the string perturbation theory to be valid.

Since string ball is lighter than black hole, more string balls are expected to be produced at
LHC than black holes. The Hagedorn temperature of a string ball is given by
\bea
T_{SB}=\frac{M_s}{\sqrt{8} \pi}.
\label{tsb}
\eea
The Hawking temperature of a black hole is
\be
T_{BH}=\frac{n+1}{4\pi R_{BH}} ~=~
\frac{n+1}{4 \sqrt{\pi}}~
M_P [\frac{M_P}{M_{BH}}\frac{n+2}{8\Gamma(\frac{n+3}{2})}]^{\frac{1}{1+n}}\,,
\label{tbh}
\ee
where $R_{BH}$ is the Schwarzschild radius of the black hole, $M_{BH}$ is the black hole mass
and $n$ is the number of extra dimensions.
Since these temperatures are very high at LHC ($\sim$ hundreds of GeV) we expect more massive particles
($M \sim 3 T$) to be produced at CERN LHC from string balls/black holes.

Note that in our calculation we have used the scheme advocated in \cite{dimo}, {\it i.e.,} when
Schwarzschild radius $R_s$ shrinks below the string length $l_s$ ($R_s <l_s$) black holes can
not exist as semi-classical objects with a well defined Hawking temperature, because for
$T>M_s$ the number of thermally-accessible species diverges exponentially and one is bound by the
Hagedorn effect. As shown in \cite{dvali} this naive argument is false and it alone can not
prevent semi-classical black holes from continuing existence in the sub-$l_s$ domain. The reason
is that, although the number of states is indeed exponentially large, only few of them are
effectively produced in black hole evaporation. The effective number of emitted string resonances
is maximum:
\bea
N_{eff} \sim \frac{1}{g_s^2}
\label{sth}
\eea
where $g_s$ is the string coupling constant. For a detailed calculation of this suppression, see \cite{dvali}.
From this point of view it will be interesting to implement this calculation in the study of black hole
production at LHC. One might expect to see more standard model particle production
from black holes because the temperature of the black hole in this sub-$l_s$ domain becomes large. This may
enhance the chance for black hole detection at LHC. Hence the calculation done in \cite{dvali} may have a
significant impact on the search for black hole production at LHC.

The main channel for direct Higgs production at LHC at $\sqrt{s}$ = 14 TeV is via parton (gluon) fusion processes.
The differential cross section for Higgs production in pp collisions at LHC is given by \cite{rsvn,smith}
\begin{eqnarray}
\label{eqn4.1}
d\sigma=\sum_{a,b=q,\bar q,g} \int dx_1 \int dx_2 ~f_{a/P}(x_1,\mu^2)~f_{b/P}(x_2,\mu^2 ) d{\hat \sigma}_{ab}
\end{eqnarray}
where $d{\hat \sigma}_{ab}$ is differential cross section for Higgs production in parton fusion processes. There are
contributions from WW/ZZ fusion processes to inclusive Higgs production at LHC but their contributions will be small,
see below, where we also make a comparison with the other backgrounds to standard model Higgs production at LHC.

As discussed in \cite{gl,yu,smithhiggs} there is another mode of Higgs production which is via Hawking's radiation
of black holes at LHC. The decay of Higgs which are produced from black holes is studied in \cite{arif}.
In this paper we make a detail analysis of Higgs production and decay from string balls at LHC. We make a
comparison of this with the Higgs production and decay from from black holes and with the standard model Higgs.
For light Higgs decay we mainly focus on diphoton decay channel with Higgs mass
$M_H$ = 120 GeV. For the heavy Higgs decay we focus on $ZZ$ golden decay channel with Higgs
mass $M_H$ = 200 GeV.

The differential cross section for Higgs with mass $M_H$, momentum $\vec{p}$ and energy
$E =\sqrt{{\vec p}^2+M_H^2}$ from a black hole of temperature $T_{BH}$ is given by \cite{arif}
\bea
\frac{Ed\sigma_{\rm Higgs}}{d^3p}
= \frac{1}{(2\pi)^3s}{\sum}_{ab}~\int_{\frac{M_s^2}{g_s^4}}^s dM^2~\int \frac{dx_a}{x_a} ~f_{a/p}(x_a, \mu^2)~f_{b/p}(\frac{M^2}{sx_a}, \mu^2)~\hat{\sigma}^{ab}(M)~\frac{c_n \sigma_n \gamma \tau_{BH} p^\mu u_\mu }{(e^{\frac{p^\mu u_\mu}{T_{BH}}} - 1)}\,, \nonumber \\
\label{bksh1}
\eea
where $M_s$ is the string mass scale and $g_s$ is the string coupling which is less than $1$ for the string perturbation theory to be valid,
see eq. (\ref{scale}). The flow velocity is $u^\mu$ and $\gamma$ is the Lorentz boost factor with
\bea
\gamma {\vec v}_{BH} = (0,0,\frac{(x_1-x_2) \sqrt{s}}{2M_{BH}}).
\label{4v}
\eea
We take the grey body factor $\sigma_n$ in the geometrical approximation \cite{grey}
\begin{equation}
\sigma_n=\pi\left(\frac{n+3}{2}\right)^{2/(n+1)}\frac{n+3}{n+1}R^2_{\rm BH}.
\end{equation}
The partonic level cross section of the black hole can be approximated with the pure geometrical form
$\hat{\sigma}^{ab}(M_{BH}) \sim \pi R^2_{BH}$. The multiplicity factor $c_n=1$ for Higgs
because there is no color or spin degrees of freedom \cite{gram}.
The decay time of the black hole is given by \cite{arif}
\bea
\tau_{BH}= \frac{1}{M_P} (\frac{M_{BH}}{M_P})^{\frac{n+3}{n+1}}.
\label{tb}
\eea

Similarly, the differential cross section for Higgs with mass $M_H$, momentum $\vec{p}$ and energy
$E =\sqrt{{\vec p}^2+M_H^2}$ from a string ball of temperature $T_{SB}$ is given by
\bea
&& \frac{Ed\sigma_{\rm Higgs}}{d^3p}
= \frac{1}{(2\pi)^3s}{\sum}_{ab}~\int_{M^2_{s}}^{\frac{M^2_s}{g^4_s}}~dM^2~
\int \frac{dx_a}{x_a} ~f_{a/p}(x_a, \mu^2)~f_{b/p}(\frac{M^2}{sx_a}, \mu^2)~\hat{\sigma}^{ab}(M)~\frac{A_n c_n \gamma \tau_{SB} p^\mu u_\mu}{(e^{\frac{p^\mu u_\mu}{T_{SB}}} - 1)}\,, \nonumber \\
\label{bksh}
\eea
where $A_n$ is the $d(=n+3)$ dimensional area factor \cite{bv,nayaks}. We will use the number of extra dimensions
$n=6$ in our calculation. The partonic level string ball production cross section is given by \cite{dimo}
\bea
&& {\hat \sigma}(M_{SB}) = \frac{1}{M^2_s},~~~~~~~~~~~~~~~~~~~~~~~~\frac{M_s}{g_s}<M_{SB}<\frac{M_s}{g^2_s} \nonumber \\
&& {\hat \sigma}(M_{SB}) = \frac{g^2_sM^2_{SB}}{M^4_s},~~~~~~~~~~~~~~~~~~~~~~~~M_s<M_{SB}<\frac{M_s}{g_s}.
\label{ssb}
\eea

We consider diphoton and $ZZ$ decay channel of Higgs at LHC when the Higgs is produced from string balls and
black holes and make a comparison with the diphoton and $ZZ$ produced from standard model Higgs and with the direct
diphoton and $ZZ$ production from standard model processes. The branching ratio of Higgs decay is function of Higgs
mass \cite{hdecay}. As the temperature of the string ball (black hole) is very large at LHC there is not much
difference in the Higgs production cross section from string balls (black holes) if the Higgs mass $M_H$ is increased
from 100 to 300 GeV.

As mentioned above the string coupling $g_s$ should be less than 1 for the string perturbation theory
to be valid, see eq. (\ref{scale}). Hence we will use \cite{dimo,canada}
\bea
g^2_s=\frac{1}{3}
\eea
in our calculation. Note that, recently, the string mass scale $M_s$=1 TeV is ruled out via dijet
measurement by CMS collaboration at LHC \cite{cmslhc}. The new lower bound on the string mass scale
from this dijet measurement at LHC is $M_s$=2.5 TeV. Hence we will use string mass scale $M_s$=1 TeV
and 2.5 TeV respectively in our calculation.

We present the results of $\frac{d\sigma}{dp_T}$ of diphoton and $ZZ$ pair production
from Higgs from string balls (black holes) and from Higgs in parton fusion processes at NLO. We also
compare these with the $\frac{d\sigma}{dp_T}$ of diphoton and $ZZ$ pairs directly produced from parton
fusion processes at LHC \cite{qcddph,qcdzz}. We also compute diphoton cross section from Higgs decay
from black holes and string balls at LHC. For the diphoton channel we consider light Higgs with mass
$M_H$ = 120 GeV and for $ZZ$ decay channel we consider heavy Higgs with mass $M_H$ = 200 GeV.

String theory at LHC is also studied via string reggge excitations \cite{han,all}.  Since the matrix element for Higgs
production in parton-parton collisions via string regge excitations is not available we compute $\frac{d\sigma}{dp_T}$ of
photon production via string regge excitations at LHC and make a comparison with that from strings balls.

The $\frac{d\sigma}{dp_T}$ for photon production in parton-parton collisions via string regge
excitation scenario can be calculated by using the matrix elements from the processes
$q{\bar q} \rightarrow \gamma \gamma$ and $gg \rightarrow \gamma \gamma$. Although gluon-gluon fusion process
contribution is small at Tevatron they are not small at LHC. The tree-level four-particle open-string amplitude
for the process $q{\bar q} \rightarrow \gamma \gamma$ is given by \cite{han}
\bea
{\cal A}_{string}(q_{\alpha} {\bar q}_{\beta} \rightarrow \gamma_{\alpha} \gamma_{\beta}) = 2ie^2 Q_q^2\sqrt{\frac{{\hat t}}{{\hat u}}} S({\hat t},{\hat u})+ie^2T\frac{1}{{\hat s}}\sqrt{\frac{{\hat t}}{{\hat u}}}
f({\hat s},{\hat t},{\hat u})
\label{qqa}
\eea
where $\alpha,\beta$ stand for helicities, the Chan-Paton factor $T=T_{1234}=T_{1243}$ and
\bea
&& ~S({\hat s},{\hat t})=\frac{\Gamma[1-\frac{{\hat s}}{M_s^2}]~~\Gamma[1-\frac{{\hat t}}{M_s^2}]}{\Gamma[1-\frac{{\hat s}}{M_s^2}-\frac{{\hat t}}{M_s^2}]}, \nonumber \\
&&~f({\hat s},{\hat t},{\hat u})={\hat u}S({\hat s},{\hat t})+{\hat s}S({\hat t},{\hat u})+{\hat t}S({\hat u},{\hat s}).
\label{fstu}
\eea
The open-string amplitude for $gg \rightarrow \gamma \gamma$ process is given by
\bea
{\cal A}_{string}(g_{\alpha} g_{\beta} \rightarrow \gamma_{\alpha} \gamma_{\beta}) = ie^2 T\frac{t}{{\hat u}{\hat s}}f({\hat s},{\hat t},{\hat u})
\label{gga}
\eea
where the Chan-Paton factor $T=T_{1234}=T_{1324}=T_{1243}$. We will take $0\le T \le 4$ \cite{han}.

Note that since we discuss diphoton production from Higgs, we consider the invariant mass
range $110 \le M_{\gamma \gamma} \le 140$ GeV of the diphoton in this paper. This invariant mass
of diphoton is small comparison to the TeV scale string mass scale $M_s$. Hence, in order to
compare low $p_T$ photon production from tree-level four-particle open-string amplitude with that
from from string balls it is sufficient to consider the low energy limit (${\hat s} < M_s^2$) of the
tree-level four-particle open-string amplitude. Note that for high $p_T$ ($\sim$ TeV) photon production
one will observe resonances in the limit ${\hat s}=n M_s^2$ where $n$=1,2,...etc.

In the low-energy limit (${\hat s} < M_s^2$), the matrix element square from the open string amplitudes
in the quark-antiquark process is given by \cite{han}
\bea
&& |M(q{\bar q} \rightarrow \gamma \gamma )|^2 = \frac{1}{12}[16 e^4 Q_q^4 \frac{1+{\rm cos}^2\theta}{1-{\rm cos}^2\theta}-4e^4
\frac{\pi^2}{3} (Q_q^2+\frac{3}{2}T) Q_q^2\frac{{\hat s}^2}{M_s^4}(1+{\rm cos}^2\theta)\nonumber \\
&&+\frac{e^4}{4}\frac{\pi^4}{9}(Q_q^2+\frac{3}{2}T)^2\frac{{\hat s}^4}{M_s^8} (1-{\rm cos}^4\theta)]
\label{mm}
\eea
and in the gluon-gluon process
\bea
&& |M(gg \rightarrow \gamma \gamma )|^2 = \frac{g^4\pi^4T^2}{64}\frac{{\hat s}^4}{8M_s^8} (1+6{\rm cos}^2\theta+{\rm cos}^4\theta)
\label{mm1}
\eea
where
\bea
-\frac{1}{2}(1+{\rm cos}\theta)=\frac{{\hat u}}{{\hat s}},~~~~~~~~~~~~~~~~~~~-\frac{1}{2}(1-{\rm cos}\theta)=\frac{{\hat t}}{{\hat s}}.
\label{th}
\eea
The $\frac{d\sigma}{dp_T}$ of photon production is given by
\bea
\frac{d\sigma}{dp_T} = \sum_{a,b}~\frac{p_T}{{8 \pi s}}~\int dy \int dy_2 ~\frac{1}{{\hat s}}~f_a(x_1, Q^2)~ f_b(x_2,Q^2)~\times~
|M(ab \rightarrow \gamma \gamma )|^2
\label{dsdpt}
\eea
where $a,b=q,{\bar q}, g$, the matrix element square
$|M(ab \rightarrow \gamma \gamma )|^2$ is given by eqs. (\ref{mm}) and (\ref{mm1}) and
\bea
x_1=\frac{p_T}{\sqrt{s}} [e^{y}+e^{y_2}],~~~~~~~~~~~~~~x_2=\frac{p_T}{\sqrt{s}} [e^{-y}+e^{-y_2}].
\label{x12}
\eea

We have used CTEQ6M sets \cite{cteq} for the parton distribution function $f(x,Q^2)$ inside
the proton at LHC.

\begin{figure}[htb]
\vspace{2pt}
\centering{{\epsfig{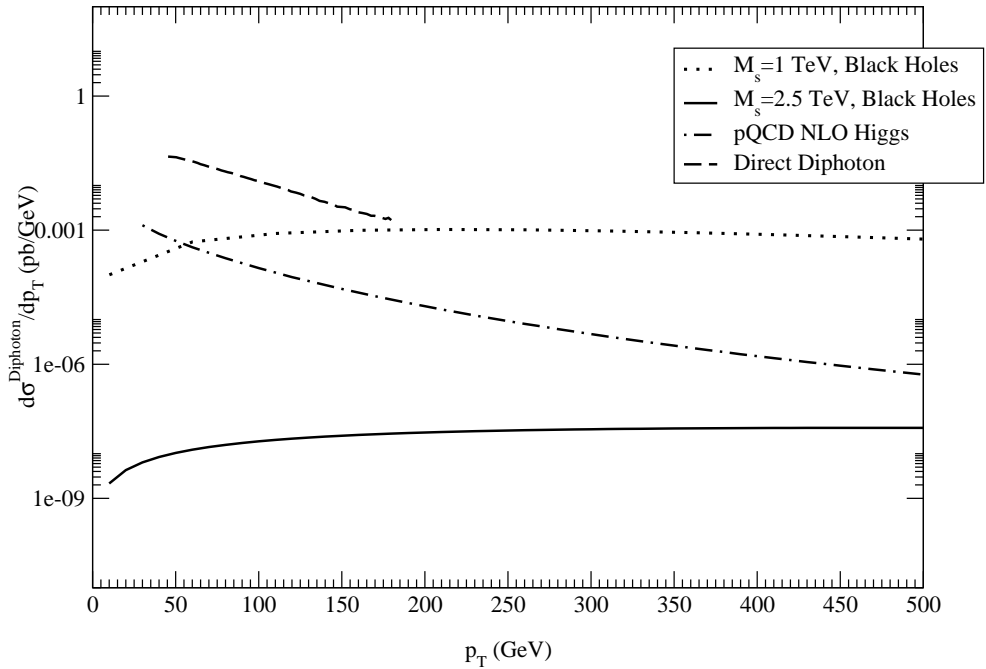}}}
\caption{$p_T$ distribution of diphoton production from Higgs from black holes at LHC and from
Higgs from parton fusion processes using pQCD at LHC.}
\label{fig1}
\end{figure}

In Fig.1 we present the results of $\frac{d\sigma}{dp_T}$ of diphoton production from Higgs from black holes
and make a comparison with the diphoton produced from Higgs from parton fusion processes at LHC and with the
$\frac{d\sigma}{dp_T}$ of diphoton produced directly from parton fusion processes at LHC \cite{qcddph}.
We have used the Higgs mass
$M_H$ = 120 GeV in our calculation for diphoton case. The dotted line is for diphoton from Higgs from black holes with
string mass scale $M_s$ = 1 TeV. The solid line is for diphoton from Higgs from black holes with
string mass scale $M_s$ = 2.5 TeV. The dot-dashed line is for diphoton from Higgs where the Higgs is produced
from parton fusion processes at LHC. The dashed line is for $\frac{d\sigma}{dp_T}$ of diphoton produced directly from parton
fusion processes at LHC \cite{qcddph}.

It can be seen from Fig.1 that for $M_s$=1 TeV we find that $\frac{d\sigma}{dp_T}$ of high $p_T$ ($\sim $ 300-500 GeV)
diphotons from Higgs from black holes at LHC can be larger than $\frac{d\sigma}{dp_T}$ of diphotons from standard model Higgs and also
can be larger than $\frac{d\sigma}{dp_T}$ of diphotons from direct pQCD processes. For $M_s$=2.5 TeV, which is the lower limit of the
string mass scale set by the recent CMS collaboration at LHC \cite{cmslhc}, the $\frac{d\sigma}{dp_T}$ of diphoton production from
Higgs from black holes is smaller than that from the standard model Higgs for $p_T$ of the diphotons up to 500 GeV.
In both the cases we find that for larger values of $p_T$, the $\frac{d\sigma}{dp_T}$ of diphotons from Higgs from black holes
increases with increasing $p_T$ whereas the $\frac{d\sigma}{dp_T}$ of diphotons from Higgs from NLO pQCD processes decreases rapidly.
Hence the $p_T$ distribution of high $p_T$ diphotons at LHC can be a good signature to study black hole production at LHC.

\begin{figure}[htb]
\vspace{2pt}
\centering{{\epsfig{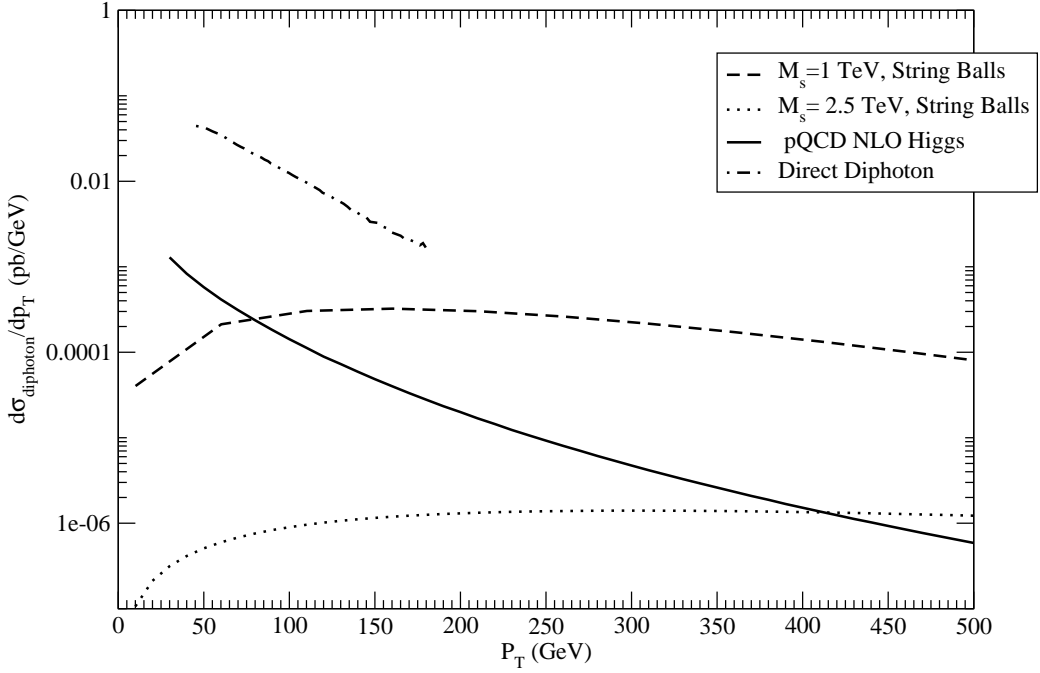}}}
\caption{ $p_T$ distribution of diphoton production from Higgs from string balls at LHC and from
Higgs from parton fusion processes using pQCD at LHC}.
\label{fig2}
\end{figure}

In Fig.2 we present the results of $\frac{d\sigma}{dp_T}$ of diphoton production from Higgs from
string balls and make a comparison with the diphoton produced from Higgs from parton fusion processes
at LHC and with $\frac{d\sigma}{dp_T}$ of diphoton produced directly from parton
fusion processes at LHC \cite{qcddph}. We have used the Higgs mass
$M_H$ = 120 GeV in our calculation for diphoton case. The dashed line is for diphoton from Higgs from string balls with string mass scale
$M_s$ = 1 TeV. The dotted line is for diphoton from Higgs from string balls with string mass scale
$M_s$ = 2.5 TeV.  The solid line is for diphoton from Higgs where the Higgs is produced from parton fusion processes at LHC.
The dot-dashed line is for $\frac{d\sigma}{dp_T}$ of diphotons produced directly from parton
fusion processes at LHC \cite{qcddph}.

It can be seen from Fig.2 that for $M_s$=2.5 TeV, which is the lower limit of the
string mass scale set by the recent CMS collaboration at LHC \cite{cmslhc}, the $\frac{d\sigma}{dp_T}$ of high $p_T$
($\ge$ 450 GeV) diphoton production from Higgs from black holes is larger than that from the standard model Higgs.
For $M_s$=1 TeV and for diphotons with $p_T \ge $ 100 GeV the $\frac{d\sigma}{dp_T}$ of diphoton
production from Higgs from string balls is larger than $\frac{d\sigma}{dp_T}$ of diphotons produced from standard model
Higgs.

Hence we find that $\frac{d\sigma}{dp_T}$ of high $p_T$ ($\ge $ 450 GeV) diphoton from Higgs
from string balls at LHC can be larger than $\frac{d\sigma}{dp_T}$ of diphotons from standard model Higgs.
We find that for larger values of $p_T$, the $\frac{d\sigma}{dp_T}$ of diphotons from Higgs from string
balls does not decrease sharply with increase in $p_T$, whereas in case of NLO pQCD processes it decreases
rapidly. On the other hand $\frac{d\sigma}{dp_T}$ of diphotons from Higgs from black holes increases with
increasing $p_T$ (see Fig. 1).

\begin{figure}[htb]
\vspace{2pt}
\centering{{\epsfig{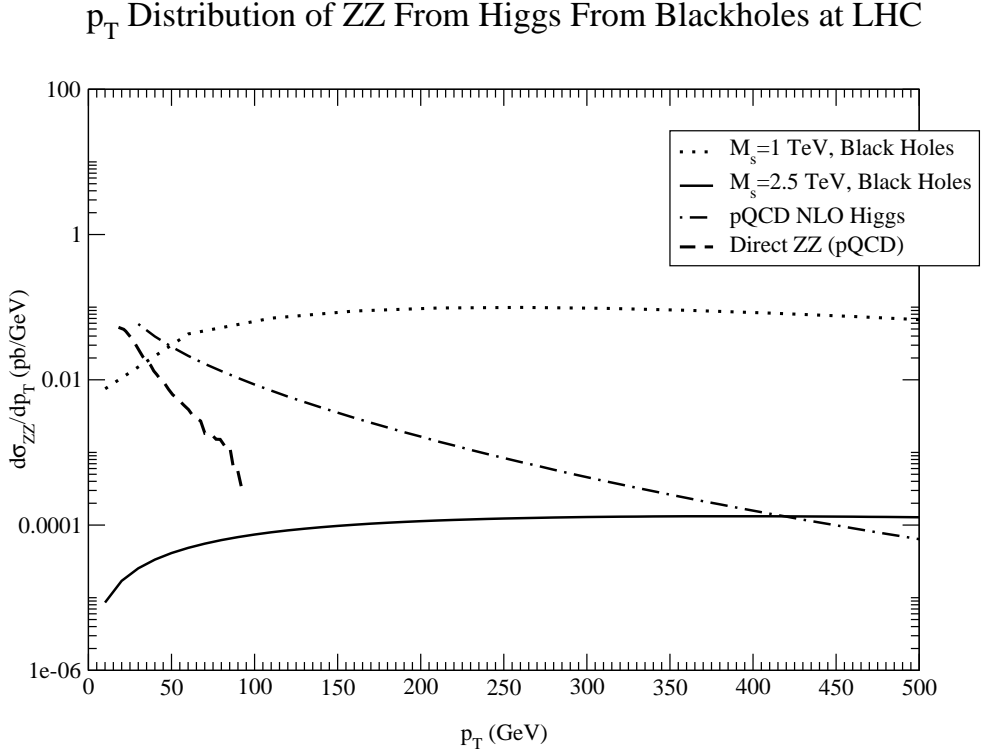}}}
\caption{$p_T$ distribution of $ZZ$ pair production from Higgs from black holes at LHC and from
Higgs from parton fusion processes using pQCD at LHC.}
\label{fig3}
\end{figure}

In Fig.3 we present the results of $\frac{d\sigma}{dp_T}$ of $ZZ$ pairs produced from Higgs from
black holes and make a comparison with the $ZZ$ produced from Higgs from parton fusion
processes at LHC and with $\frac{d\sigma}{dp_T}$ of $ZZ$ pairs produced directly from parton
fusion processes at LHC \cite{qcdzz}.
We have used the Higgs mass $M_H$ = 200 GeV in our calculation for $ZZ$ case.
The dotted line is for $ZZ$ from Higgs from black holes with string mass scale $M_s$ = 1 TeV.
The solid line is for $ZZ$ from Higgs from black holes with string mass scale $M_s$ = 2.5 TeV.
The dot-dashed line is for $ZZ$ from Higgs where the Higgs is produced
from parton fusion processes at LHC. The dashed line is the $\frac{d\sigma}{dp_T}$ of $ZZ$ pairs
produced directly from parton fusion processes at LHC \cite{qcdzz}.

It can be seen from Fig.3 that for $M_s$=2.5 TeV, which is the lower limit of the
string mass scale set by the recent CMS collaboration at LHC \cite{cmslhc}, the $\frac{d\sigma}{dp_T}$
of high $p_T$ ($\ge$ 400 GeV) $ZZ$ pair production from Higgs from black holes is larger than that from the standard model Higgs.
Although the $\frac{d\sigma}{dp_T}$ of $ZZ$ pairs produced from direct parton fusion processes is not
available at very high $p_T$ the trend of the plot suggests that the $\frac{d\sigma}{dp_T}$ of $ZZ$ pairs produced from Higgs
from black holes may be larger than that from the $ZZ$ pairs produced from direct parton fusion processes for $p_T$ larger
than 150 GeV. For $M_s$=1 TeV we find that the $\frac{d\sigma}{dp_T}$ of $ZZ$ pairs with $p_T \ge $ 70 GeV
produced from Higgs from black holes is larger than $\frac{d\sigma}{dp_T}$ of $ZZ$
pairs produced from the standard model Higgs.

Hence we find that $\frac{d\sigma}{dp_T}$ of $ZZ$ pairs produced
from Higgs from black holes at LHC can be larger than $\frac{d\sigma}{dp_T}$ of $ZZ$ pairs produced from standard model Higgs and also
can be larger than $\frac{d\sigma}{dp_T}$ of $ZZ$ pairs produced from direct parton fusion processes. We find that for larger values
of $p_T$, the $\frac{d\sigma}{dp_T}$ of $ZZ$ pairs produced from Higgs from black holes increases with increasing $p_T$, whereas in case
of NLO pQCD processes it decreases rapidly.

\begin{figure}[htb]
\vspace{2pt}
\centering{{\epsfig{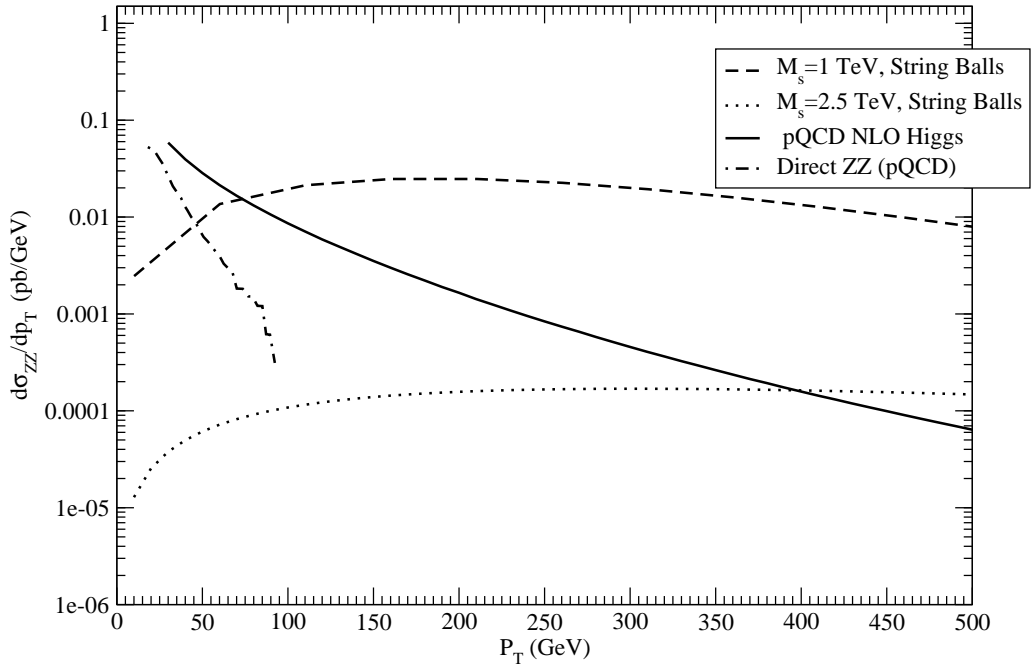}}}
\caption{$p_T$ distribution of $ZZ$ pair production from Higgs from string balls at LHC and from
Higgs from parton fusion processes using pQCD at LHC.}
\label{fig4}
\end{figure}

\begin{figure}[htb]
\vspace{2pt}
\centering{{\epsfig{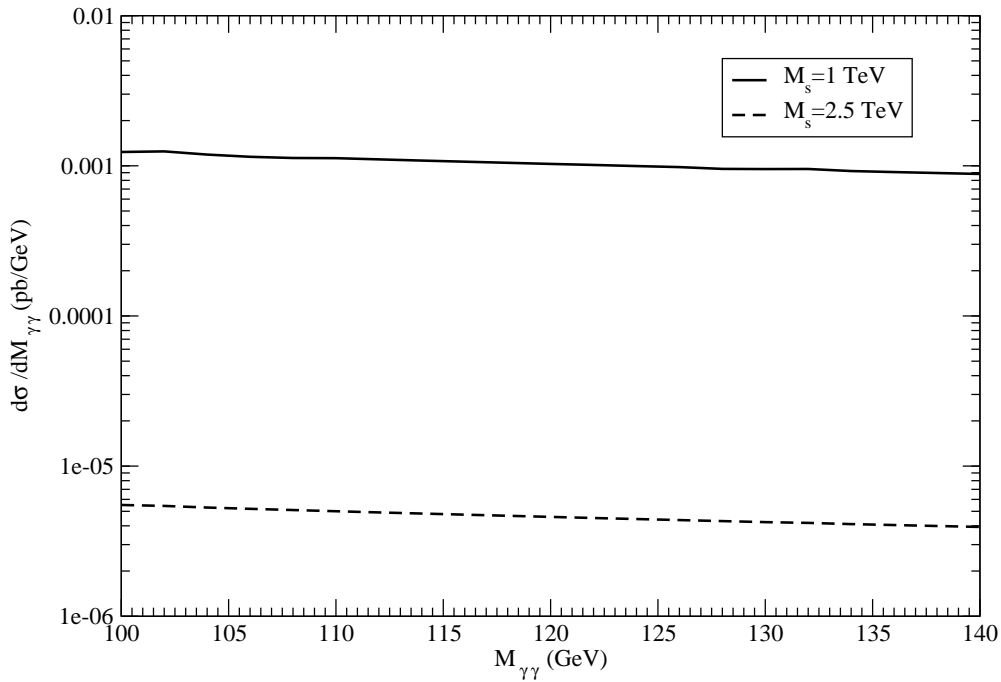}}}
\caption{Invariant mass spectrum of diphoton from Higgs from string balls at LHC.}
\label{fig5}
\end{figure}

In Fig.4 we present the results of $\frac{d\sigma}{dp_T}$ of $ZZ$ pairs produced from Higgs from
string balls and make a comparison with the $ZZ$ produced from Higgs from parton fusion processes
at LHC and with the
$\frac{d\sigma}{dp_T}$ of $ZZ$ pairs produced directly from parton fusion processes at LHC \cite{qcdzz}.
We have used the Higgs mass $M_H$ = 200 GeV in our calculation for the $ZZ$ case.
The dashed line is for $ZZ$ from Higgs from string balls with string mass scale $M_s$ = 1 TeV.
The dotted line is for $ZZ$ from Higgs from string balls with string mass scale $M_s$ = 2.5 TeV.
The solid line is for $ZZ$ from Higgs where the Higgs is produced from parton fusion
processes at LHC. The dot-dashed line is the $\frac{d\sigma}{dp_T}$ of $ZZ$ pairs produced directly from parton
fusion processes at LHC \cite{qcdzz}.

It can be seen from Fig.4 that for $M_s$=2.5 TeV, which is the lower limit of the
string mass scale set by the recent CMS collaboration at LHC \cite{cmslhc}, the $\frac{d\sigma}{dp_T}$ of high $p_T$
($\ge $ 400 GeV) $ZZ$ pair production from Higgs from string balls is larger than that from the standard model Higgs.
Although the $\frac{d\sigma}{dp_T}$ of $ZZ$ pairs produced from direct parton fusion processes is not
available at very high $p_T$ the trend of the plot suggests that the $\frac{d\sigma}{dp_T}$ of $ZZ$ pairs produced from Higgs
from string balls may be larger than that from the $ZZ$ pairs produced from direct parton fusion processes for $p_T$ larger
than 150 GeV. For $M_s$=1 TeV and $ZZ$ pairs with $p_T \ge $ 100 GeV, the $\frac{d\sigma}{dp_T}$
of $ZZ$ pairs produced from Higgs from string balls is larger than $\frac{d\sigma}{dp_T}$ of $ZZ$ pairs produced from
standard model Higgs and also is larger than $\frac{d\sigma}{dp_T}$ of $ZZ$ pairs produced from direct parton fusion processes.

Hence we find that for $p_T\ge$ 400 GeV, the $\frac{d\sigma}{dp_T}$ of $ZZ$ pairs produced from Higgs
from string balls at LHC can be larger than $\frac{d\sigma}{dp_T}$ of $ZZ$ pairs produced from standard model Higgs and also can be larger than
$\frac{d\sigma}{dp_T}$ of $ZZ$ pairs produced from direct parton fusion processes.
We find that for larger values of $p_T$, the $\frac{d\sigma}{dp_T}$ of $ZZ$ pairs from Higgs from string balls
does not decrease sharply with increase in $p_T$, whereas in case of NLO pQCD processes it decreases rapidly. On the other hand
$\frac{d\sigma}{dp_T}$ of $ZZ$ from Higgs from black holes increases with increasing $p_T$ (see Fig. 3).

\begin{figure}[htb]
\vspace{2pt}
\centering{{\epsfig{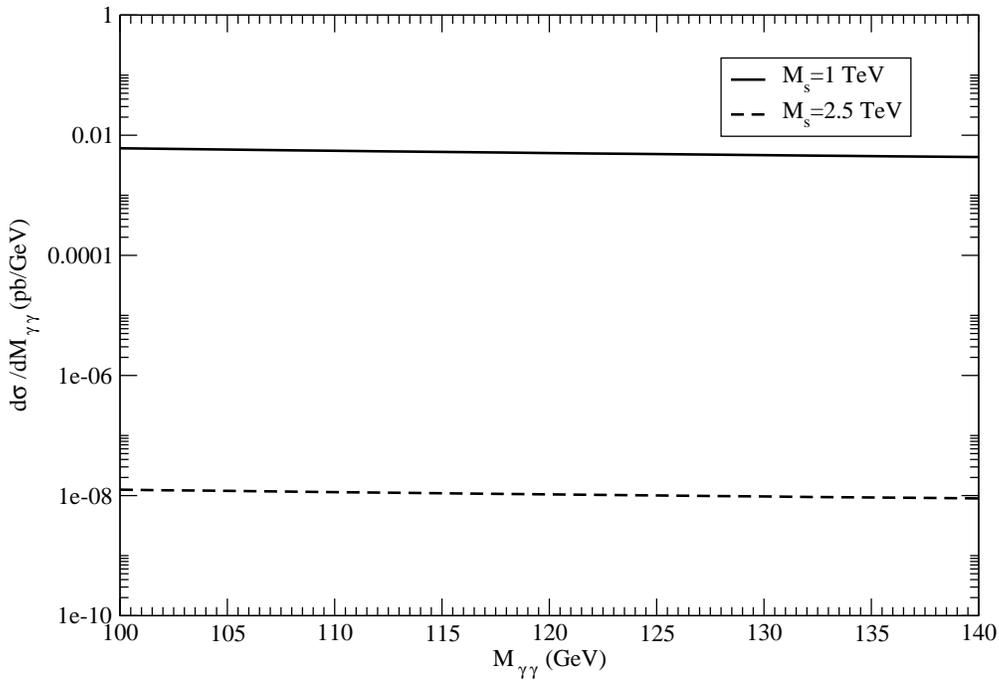}}}
\caption{Invariant mass spectrum of diphoton from Higgs from black holes at LHC.}
\label{fig6}
\end{figure}

These results suggest that if string mass scale $M_s \sim$ 2.5 TeV, which is the lower limit of the
string mass scale set by the recent CMS collaboration at LHC \cite{cmslhc}, then $\frac{d\sigma}{dp_T}$ of high $p_T$ diphotons and $ZZ$ pairs
can be used to distinguish between Higgs production from: 1) black holes, 2) string balls and 3) parton fusion processes at CERN LHC.

In the mass range $110 < m_H < 140$ GeV
the Higgs boson is expected to decay into two photons with a branching fraction large enough to render the
search feasible at LHC \cite{higgsdph}. For this reason we present our result of $\frac{d\sigma}{dM_{\gamma \gamma}}$
vs $M_{\gamma \gamma}$ in the diphoton invariant mass range $110 < M_{\gamma \gamma} <140$ GeV. Note that we have used
the branching ratio of Higgs decay which is a function of Higgs mass \cite{hdecay}. The applied transverse momentum cuts
are from ATLAS \cite{atlas,atlas2}. For transverse momentum cuts at CMS, see \cite{cmshiggs}.

In Fig.5 we present $\frac{d\sigma}{dM_{\gamma \gamma}}$ of diphoton from Higgs from string balls at LHC
as a function of diphoton invariant mass. The solid line is with string mass scale $M_s$ = 1 TeV and the
dashed line is with $M_s$=2.5 TeV.  For the corresponding results from background processes and from standard model
Higgs, see \cite{atlas,atlas2}. Similarly, in Fig.6 we present $\frac{d\sigma}{dM_{\gamma \gamma}}$ of diphoton from Higgs from black holes at LHC
as a function of diphoton invariant mass. The solid line is with string mass scale $M_s$ = 1 TeV and the
dashed line is with $M_s$=2.5 TeV. For the corresponding results from background processes and from standard model
Higgs, see \cite{atlas,atlas2}. Our results suggest that $\frac{d\sigma}{dM_{\gamma \gamma}}$ of diphoton
from Higgs from string balls and black holes at LHC does not decrease rapidly as the diphoton invariant mass $M_{\gamma \gamma}$
increases.

In table-I we present the results for diphoton cross sections at LHC from different signal ($M_H$=120 GeV)
and background processes. For applied transverse momentum cuts and diphoton mass window at ATLAS,
see \cite{atlas2}.

It can be seen from table-I that the diphoton cross section from Higgs from string balls at LHC is
larger than that from standard model Higgs when string mass scale $M_s$=1 TeV.
It is also comparable to the diphoton cross section from background processes.
The result is more striking from mini-black holes. For example, if string mass scale $M_s$=1 TeV
then the diphoton cross section from Higgs from black holes is larger than that
from standard model Higgs plus background processes. However for string mass scale $M_s \sim$ 2.5 TeV,
which is the lower limit of the string mass scale set by the recent CMS collaboration at LHC \cite{cmslhc},
the diphoton cross section from Higgs from string balls at LHC is smaller than that from standard model Higgs
and from background processes.

\begin{table}[tb]
\begin{center}
\begin{tabular}{|c||c|c|}
	\hline
\textbf{ $\gamma \gamma$} at LHC from:  &  Cross Section (fb)    &   $M_s$  (TeV)    \\
	\hline \hline
Background processes &562 \cite{atlas2} &    \\
\hline
Standard model Higgs  & 25.3 \cite{atlas2}   &    \\
\hline
Higgs from string balls  & 110 & 1 \\
\hline
Higgs from string balls  & 0.56 & 2.5 \\
	\hline
Higgs from black holes  & 616 &  1  \\
\hline
Higgs from black holes  & 0.0015 &  2.5  \\
\hline
\end{tabular}
\end{center}
\caption{Diphoton cross section from different signal (Higgs) and background processes at LHC. For
transverse momentum cuts and diphoton invariant mass window, see \cite{atlas2}.}
\end{table}

Since the matrix element for Higgs production in parton-parton collisions via string regge excitations
is not available we compute $\frac{d\sigma}{dp_T}$ of photon production from string regge excitations at LHC and make a
comparison with that from strings balls.

\begin{figure}[htb]
\vspace{2pt}
\centering{{\epsfig{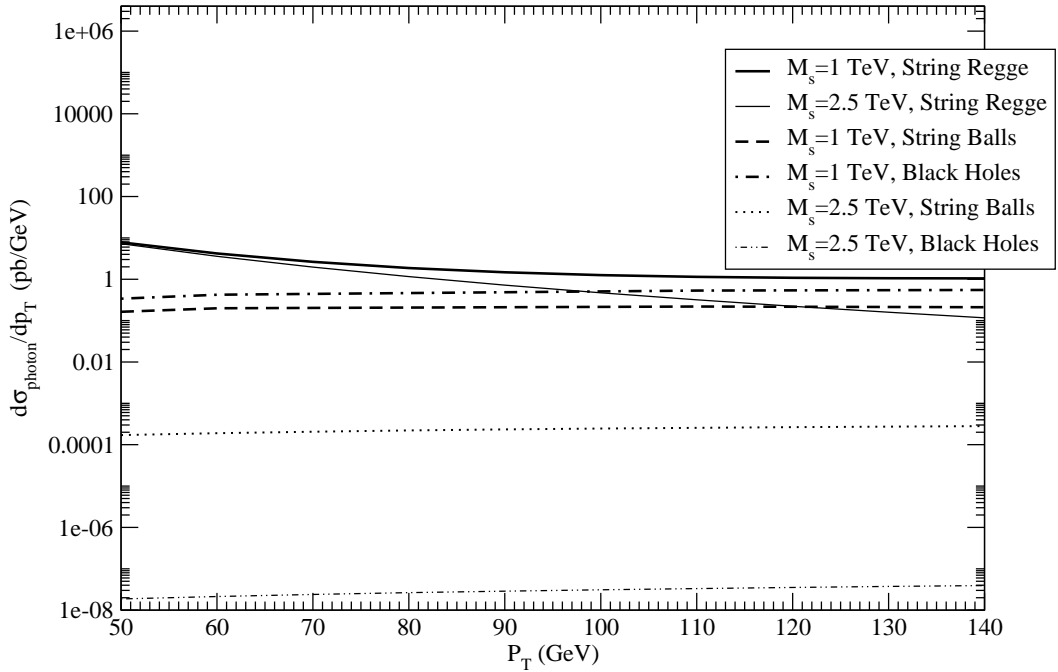}}}
\caption{Photon production from string regge excitations vs string balls at LHC. }
\label{fig7}
\end{figure}

In Fig.7 we present $\frac{d\sigma}{dp_T}$ of photon production in parton-parton collisions
at LHC from the tree level four-particle open-string amplitudes via string regge excitations
and make a comparison with
that produced from string balls. The thick solid line is for $\frac{d\sigma}{dp_T}$
of photon production from the tree level four-particle open-string amplitudes via string regge
excitations with Chan-Paton factor
$T=1$ and string mass scale $M_s$=1 TeV. The thin solid line is for $\frac{d\sigma}{dp_T}$
of photon production from the tree level four-particle open-string amplitudes via string regge
excitations with Chan-Paton factor
$T=1$ and string mass scale $M_s$=2.5 TeV.
Note that in this low $p_T$ region the results do not change significantly if we choose
the Chan-Paton factor $T=4$. The dashed line is for $\frac{d\sigma}{dp_T}$ of
photon production from string balls at LHC with string mass scale $M_s$=1 TeV.
The dotted line is for $\frac{d\sigma}{dp_T}$ of
photon production from string balls at LHC with string mass scale $M_s$=2.5 TeV.
The dot-dashed line is for $\frac{d\sigma}{dp_T}$ of photon
produced from black holes at LHC with $M_s$=1 TeV.
The dot-dot-dashed line is for $\frac{d\sigma}{dp_T}$ of photon
produced from black holes at LHC with $M_s$=2.5 TeV.

It can be seen from Fig.7 that for $M_s$=2.5 TeV, which is the lower limit of the
string mass scale set by the recent CMS collaboration at LHC \cite{cmslhc}, the $\frac{d\sigma}{dp_T}$
of photon production from string regge excitations is larger than that from string balls and black holes at LHC.

In conclusion we have studied Higgs production and decay from TeV scale string balls in $pp$ collisions at $\sqrt{s}$ = 14 TeV at LHC.
We have presented the results of total cross section of diphotons, invariant mass distribution of the diphotons and $p_T$
distribution of the diphotons and $ZZ$ pairs from Higgs from string balls at LHC.
We have found that the invariant mass distribution of diphotons from Higgs from string balls
is not very sensitive to the increase in diphoton invariant mass.
We have found that for string mass scale $M_s$=2.5 TeV, which is the lower limit of the
string mass scale set by the recent CMS collaboration at LHC, the
$\frac{d\sigma}{dp_T}$ of high $p_T$ ($\ge$ 450 GeV) diphotons and $ZZ$ pairs produced from Higgs from string balls
is larger than that from standard model Higgs. Hence in the
absence of black hole production at LHC an enhancement of high $p_T$ diphotons and $ZZ$
pairs at LHC can be useful signatures for string theory at LHC. Since the matrix element for Higgs production in parton-parton
collisions via string regge excitations is not available we have computed $\frac{d\sigma}{dp_T}$ of photon production from string regge excitations
and make a comparison with that from string balls at LHC. We have found that for string mass scale
$M_s$ = 2.5 TeV the $\frac{d\sigma_{photon}}{dp_T}$ from string
regge excitations is larger than that from string balls and black holes at LHC.

\acknowledgments

This work was supported in part by Department of Energy under contracts DE-FG02-91ER40664, DE-FG02-04ER41319
and DE-FG02-04ER41298.

\end{document}